\documentclass[12pt]{article}
\usepackage{amsfonts}
\usepackage{amssymb}
\usepackage{graphics,psboxit,amsmath, epsfig}
\usepackage{psfrag}

\def\hybrid{\topmargin -20pt    \oddsidemargin 0pt
        \headheight 0pt \headsep 0pt
        \textwidth 6.35in       
        \textheight 9.25in       
        \marginparwidth .875in
        \parskip 5pt plus 1pt   \jot = 1.5ex}

\hybrid

\def\baselinestretch{1.2}

\catcode`\@=11

\def\marginnote#1{}
\newcount\hour
\newcount\minute
\newtoks\amorpm
\hour=\time\divide\hour by60
\minute=\time{\multiply\hour by60 \global\advance\minute by-\hour}
\edef\standardtime{{\ifnum\hour<12 \global\amorpm={am}%
        \else\global\amorpm={pm}\advance\hour by-12 \fi
        \ifnum\hour=0 \hour=12 \fi
        \number\hour:\ifnum\minute<10 0\fi\number\minute\the\amorpm}}
\edef\militarytime{\number\hour:\ifnum\minute<10 0\fi\number\minute}

\def\draftlabel#1{{\@bsphack\if@filesw {\let\thepage\relax
   \xdef\@gtempa{\write\@auxout{\string
      \newlabel{#1}{{\@currentlabel}{\thepage}}}}}\@gtempa
   \if@nobreak \ifvmode\nobreak\fi\fi\fi\@esphack}
        \gdef\@eqnlabel{#1}}
\def\@eqnlabel{}
\def\@vacuum{}
\def\draftmarginnote#1{\marginpar{\raggedright\scriptsize\tt#1}}

\def\draft{\oddsidemargin -.5truein
        \def\@oddfoot{\sl preliminary draft \hfil
        \rm\thepage\hfil\sl\today\quad\militarytime}
        \let\@evenfoot\@oddfoot \overfullrule 3pt
        \let\label=\draftlabel
        \let\marginnote=\draftmarginnote
   \def\@eqnnum{(\theequation)\rlap{\kern\marginparsep\tt\@eqnlabel}%
\global\let\@eqnlabel\@vacuum}  }

\def\preprint{\twocolumn\sloppy\flushbottom\parindent 2em
        \leftmargini 2em\leftmarginv .5em\leftmarginvi .5em
        \oddsidemargin -.5in    \evensidemargin -.5in
        \columnsep .4in \footheight 0pt
        \textwidth 10.in        \topmargin  -.4in
        \headheight 12pt \topskip .4in
        \textheight 6.9in \footskip 0pt
        \def\@oddhead{\thepage\hfil\addtocounter{page}{1}\thepage}
        \let\@evenhead\@oddhead \def\@oddfoot{} \def\@evenfoot{} }

\def\numberbysection{\@addtoreset{equation}{section}
        \def\theequation{\thesection.\arabic{equation}}}

\def\underline#1{\relax\ifmmode\@@underline#1\else
        $\@@underline{\hbox{#1}}$\relax\fi}

\def\titlepage{\@restonecolfalse\if@twocolumn\@restonecoltrue\onecolumn
     \else \newpage \fi \thispagestyle{empty}\c@page\z@
        \def\thefootnote{\fnsymbol{footnote}} }

\def\endtitlepage{\if@restonecol\twocolumn \else \newpage \fi
        \def\thefootnote{\arabic{footnote}}
        \setcounter{footnote}{0}}  

\catcode`@=12
\relax

\def\figcap{\section*{Figure Captions\markboth
        {FIGURECAPTIONS}{FIGURECAPTIONS}}\list
        {Figure \arabic{enumi}:\hfill}{\settowidth\labelwidth{Figure
999:}
        \leftmargin\labelwidth
        \advance\leftmargin\labelsep\usecounter{enumi}}}
 \relax
\def\tablecap{\section*{Table Captions\markboth
        {TABLECAPTIONS}{TABLECAPTIONS}}\list
        {Table \arabic{enumi}:\hfill}{\settowidth\labelwidth{Table
999:}
        \leftmargin\labelwidth
        \advance\leftmargin\labelsep\usecounter{enumi}}}
 \relax
\def\reflist{\section*{References\markboth
        {REFLIST}{REFLIST}}\list
        {[\arabic{enumi}]\hfill}{\settowidth\labelwidth{[999]}
        \leftmargin\labelwidth
        \advance\leftmargin\labelsep\usecounter{enumi}}}
 \relax
\makeatletter
\newcounter{pubctr}
\def\publist{\@ifnextchar[{\@publist}{\@@publist}}
\def\@publist[#1]{\list
        {[\arabic{pubctr}]\hfill}{\settowidth\labelwidth{[999]}
        \leftmargin\labelwidth
        \advance\leftmargin\labelsep
        \@nmbrlisttrue\def\@listctr{pubctr}
        \setcounter{pubctr}{#1}\addtocounter{pubctr}{-1}}}
\def\@@publist{\list
        {[\arabic{pubctr}]\hfill}{\settowidth\labelwidth{[999]}
        \leftmargin\labelwidth
        \advance\leftmargin\labelsep
        \@nmbrlisttrue\def\@listctr{pubctr}}}
 \relax
\makeatother
\newskip\humongous \humongous=0pt plus 1000pt minus 1000pt

\newif\ifdtup

\relax

\def\be{\begin{equation}}
\def\ee{\end{equation}}
\def\ba{\begin{eqnarray}}
\def\ea{\end{eqnarray}}

\def\no{\noindent}

\def\IR{\relax{\rm I\kern-.18em R}}
\def\II{\relax{\rm 1\kern-.35em1}}

\renewcommand{\theequation}{\thesection.\arabic{equation}}
\csname @addtoreset\endcsname{equation}{section}

\def\IR{\relax{\rm I\kern-.18em R}}
\def\inv{^{\raise.15ex\hbox{${\scriptscriptstyle -}$}\kern-.05em 1}}

\begin{document}

\begin{titlepage}
\begin{center}

\hfill IFT-UAM/CSIC-08-44 \\
\vskip -.1 cm
\hfill arXiv: 0807.2339 \\

\vskip .5in

{\LARGE Magnons and BFKL}
\vskip 0.4in

{\bf C\'esar G\'omez}, {\bf Johan Gunnesson}\phantom{x}and\phantom{x}
{\bf Rafael Hern\'andez} 
\vskip 0.1in

\vskip .2in

Instituto de F\'{\i}sica Te\'orica UAM/CSIC \\
Facultad de Ciencias, C-XVI, 
Universidad Aut\'onoma de Madrid \\
Cantoblanco, 28049 Madrid, Spain\\
{\footnotesize{\tt cesar.gomez@uam.es, johan.gunnesson@uam.es, r.hernandez@uam.es}}

\end{center}

\vskip .4in

\centerline{\bf Abstract}
\vskip .1in
\no
We extract from the double logarithmic contributions to DGLAP anomalous 
dimensions for twist-two operators up to three-loops the magnon dispersion 
relation for planar ${\cal N}=4$ supersymmetric Yang-Mills. Perturbatively the magnon dispersion relation agrees with the expansion of the 
anomalous dimension for spin-one as well as with the non-collinear double 
logarithmic contributions to the BFKL anomalous dimensions analytically extended 
to negative spin. The all-loop expression for the magnon dispersion relation 
is determined by the double logarithmic resummation of the corresponding 
Bethe-Salpeter equation. A potential map relating the spin chain magnon to 
BFKL eigenfunctions in the double logarithm approximation is suggested.

\noindent

\vskip .4in
\noindent

\end{titlepage}
\vfill
\eject

\def\baselinestretch{1.2}


\baselineskip 20pt


\section{Introduction}

\no
Integrable structures in a four-dimensional quantum field theory were first 
shown to arise in the Regge limit of scattering amplitudes in the planar 
limit of QCD \cite{LipatovQCD,FaddeevKorchemsky}. In the leading logarithm approximation the reggeized 
scattering amplitudes are described by a non-compact Heisenberg magnet with 
$SL(2)$ symmetry group. 
Integrability survives as the amount of symmetry is increased, because 
supersymmetric extensions of QCD share the same non-compact sector of operators with 
covariant derivatives. In fact  
integrability extends to larger sectors of the gauge theory 
\cite{Belitsky}, up to the maximally supersymmetric 
${\cal N}=4$ Yang-Mills, which is completely integrable at one-loop 
\cite{MZ,dilatation}. There is much evidence that integrability holds 
beyond one-loop in ${\cal N}=4$, and a long-range Bethe ansatz has in fact been suggested to govern 
the spectrum of anomalous dimensions of local gauge invariant composite 
operators to all order \cite{longrange}. The proposal for a Bethe ansatz 
only applies to asymptotically long single trace operators, and does not cover 
wrapping interactions, present beyond a certain order for finite-size operators. 
For the non-compact $SL(2)$ 
sector of the ${\cal N}=4$ theory, containing twist-two operators of the form
\be
\hbox{Tr}(D^{s_1}\Phi D^{s_2}\Phi) \ ,
\ee
with $s_1+s_2=N$ the total spin, the expansion of the asymptotic 
Bethe ansatz (ABA) equations completely agrees with the 
perturbative computation of the three-loop anomalous 
dimension of twist-two operators 
\cite{KLV}. However for twist-two operators wrapping effects are already present  
beyond third loop, and the ABA fails to reproduce the four-loop 
prediction for the anomalous dimension obtained from the BFKL pomeron 
\cite{LipatovStaudacher}. The pomeron singularity corresponds to the analytic 
continuation of the spin to $N=-1$. The purpose of this letter is to explore 
the $N=1$ case, which in the spin chain picture amounts to a single magnon 
excitation. The note is organized as follows. In Section 2 the anomalous dimension 
for twist-two operators with spin-one is shown to agree, up to three loops, with 
the perturbative expansion of the dispersion relation for planar ${\cal N}=4$ 
supersymmetric Yang-Mills. In Section 3 the contribution of double 
logarithms to the analytic extension to negative spin of the anomalous dimension 
is shown to correspond to the anomalous dimension at $N=1$, and we conjecture an 
interpretation for the spin chain magnon in the BFKL picture. We conclude in 
Section 4 with some discussion on our results. 


\section{The single magnon anomalous dimension}

\no
In deep inelastic scattering (DIS) processes anomalous dimensions 
for twist-two operators control the renormalization group behaviour 
of parton distribution functions under changes of the photon resolution. Let us denote 
by $F_{a}(x,\, Q^2)$ the number of partons 
of type $a$, with transversal momentum $k^2$ smaller or equal to $Q^2$ and with a fraction $x$ of the 
longitudinal momentum of the nucleon. The meaning of $Q^2$ in DIS is the virtuality of 
the photon, $Q^2=-q^2$, and $x=Q^2/s$ is 
the Bjorken variable describing the rapidity gap between the photon and the nucleon. Denoting 
by $F_{a}(N,\, Q^2)$ the 
Mellin transform, 
\be
F_{a}(N,\, Q^2) \equiv \int_0^1 dx\, x^{N-1}F_{a}(x,\, Q^2) \ ,
\ee
the DGLAP renormalization group equation is given by \cite{DGLAP}
\be
\frac{\partial{F_{a}(N,\, Q^2)}}{\partial{\log Q^2}} =\gamma_{a,b}(N) F_{b}(N,\, Q^2) \ ,
\ee
where $\gamma_{a,b}(N)$ is the DGLAP anomalous dimension matrix, 
which coincides with the anomalous dimension of a twist-two operator. 
Since we will only be interested in scalar twist-two operators in this note 
we will write $\gamma _{\phi , \phi}(N) \equiv \gamma_2(N)$ for brevity. 
Using conventions such that
\be
g^2 = \frac {\lambda}{8 \pi^2} \ ,
\ee
where $\lambda \equiv g_{\hbox{\tiny{YM}}}^2 N$ is the 't Hooft coupling constant, the 
anomalous dimension
\be
\gamma_L(N) = \sum_{n=1}^{\infty} \gamma_{L,n}(N) g^{2n} 
\ee
is given, up to three-loops, by \cite{KLV}
\begin{eqnarray}
\gamma_{2,1} (N) \!\!& = & \!\! 4 \, S_1 \ ,  \label{eq:gamma1loop}\\
\gamma_{2,2} (N) \!\! & = & \!\! -4 \, \big( S_3 + S_{-3} -2 S_{-2,1} + 2S_1 (S_2 + S_{-2}) \big) \ , 
\label{eq:gamma2loop}\\
\gamma_{2,3} (N) \!\! & = & \!\! -8 \, \big( 2S_{-3} S_2 - S_5 -2 S_{-2} S_3 -3 S_{-5} + 24 S_{-2,1,1,1} 
+ 6 (S_{-4,1} + S_{-3,2} + S_{-2,3} ) \nonumber \\ 
&& - 12 (S_{-3,1,1} + S_{-2,1,2} + S_{-2,2,1} ) 
- (S_2 + 2S_1^2) (3S_{-3} + S_3 -2S_{-2,1}) \\ 
&& - S_1 (8S_{-4} + S_{-2}^2 + 4S_2S_{-2} 
+2S_2^2 + 3S_4 -12 S_{-3,1} -10 S_{-2,2} + 16S_{-2,1,1}) \big) \ , 
\nonumber \label{eq:gamma3loop}
\end{eqnarray}
where the harmonic sums are defined through 
\ba
S_a & \equiv & S_a(N) = \sum_{j=1}^N \frac {(\hbox{sgn}(a))^j}{j^a} \ , \\
S_{a_1,\ldots,a_n} & \equiv & S_{a_1,\ldots,a_n}(N) = \sum_{j=1}^N \frac {(\hbox{sgn}(a_1))^j}{j^a_1} 
S_{a_2,\ldots,a_n}(j) \ .
\ea

The anomalous dimension for these twist-two operators can also be obtained as the energy for 
the proposed long-range $SL(2)$ integrable spin chain with $N$ magnon excitations through 
\cite{longrange}
\be
\gamma_2(N) =\sum_{i}^N E(p_{i}) \ ,
\ee
where the dispersion relation is
\be
E(p_i) = \sqrt{1+8g^2 \sin^2 \left( \frac {p_i}{2} \right)} - 1 \ , \label{eq:dispersion}
\ee
with $\left\{ p_{i} \right\}$ the set of magnon momenta solving the Bethe ansatz 
equations. At one-loop this spin chain reduces 
to the length-two $SL(2)$ XXX$_{s=-1/2}$ Heisenberg chain 
\cite{dilatation}, and the energy for $N$ magnons 
can be exactly obtained by solving the corresponding Baxter equation 
(see for instance \cite{ES}). 
The anomalous dimension obtained for twist-two scalar operators from the 
ABA coincides with equations \eqref{eq:gamma1loop}-\eqref{eq:gamma3loop}. 
It also provides a four-loop term \cite{LipatovStaudacher}
\be
\gamma_{2,4}(N)  =  16 \, (4 S_{-7} + 6S_7 + \ldots - \zeta(3) S_1 (S_3 - S_{-3} 
+ 2 S_{-2,1})) \ \label{eq:gamma4loop},
\ee
where the entire expression is presented in table 1 of reference 
\cite{LipatovStaudacher}.

In this note we are concerned with the value of $\gamma_2 (N)$ at $N=1$. In 
QCD, $\gamma _{a,b} (1)$ is an interesting quantity, because it provides the normalisation 
of the splitting functions. In ${\cal N}=4$ Yang-Mills, however, one does 
not have such an interpretation in terms of splitting functions. 
Instead, if one were forced to prescribe a value for $\gamma(1)$, the natural choice, 
obtained from the spin chain picture, would be $\gamma(1) = 0$. The reason is that
the translation invariance imposed by the trace on gauge operators implies vanishing 
momentum on states of the corresponding spin chain. 
Therefore a single magnon state could only have zero momentum, and therefore zero 
energy. However, the true value of $\gamma (1)$, as given by the expansions 
of the anomalous dimension in terms of harmonic sums, turns out to be rather 
surprising. Plugging $N=1$ into equations \eqref{eq:gamma1loop}-\eqref{eq:gamma3loop} 
and \eqref{eq:gamma4loop} gives 
\be
\gamma_2(1)=4 g^2 - 8g^4 + 32g^6 -160g^8 + {\cal O}(g^{10}) \ . \label{eq:gamma1}
\ee
This is precisely what is obtained if one expands the dispersion 
relation \eqref{eq:dispersion} for a magnon of momentum $p=\pi$. It would 
thus seem that $\gamma_2(1)$ does not provide the energy of a physical, 
zero-momentum magnon, but rather of some sort of 
``non-physical'' $p=\pi$ magnon. 
Extrapolating to all-loops we may conjecture that
\be
\gamma_2(1) = E \left(p= \pi \right) \ ,
\ee
where $E(p)$ is given by \eqref{eq:dispersion}. 
For later use, let us write 
the expansion coefficients of $\gamma_2(1)$, 
at weak-coupling as $e(i)$. The conjecture thus simply states that 
$E(p=\pi)=\sum_{i}e(i)g^{2i}$. 
  

\subsection{Twist-$L$ and analytical continuations}
\label{sec:twist3}

\no
Considering now that $p=\pi$ is the smallest non-zero momentum that a 
magnon can have on a 
chain of length $L=2$ it is tempting to speculate that a general 
expression for the anomalous dimension of twist-$L$ operators at $N=1$ 
could be
\be
\gamma_L(1) = E \left( p=\frac{2\pi}{L} \right) \ . 
\label{eq:conjL}
\ee 
At a first glance it would however seem that the above conjecture for arbitraty 
twist-$L$ fails for twist-three. In \cite{LipatovStaudacher,Beccaria} 
the twist-three anomalous dimensions up 
to four-loops are given in terms of harmonic sums. These expressions, 
as opposed to the twist-two formulae, have two distinctive features. 
Firstly, the harmonic sums only have positive indeces, and secondly 
they are evaluated at $N/2$. Naively this last property would invalidate the 
conjecture. For example, the one-loop expression is 
\be
\gamma_{3,1} (N)  =  4 \, S_1\left( \frac{N}{2}\right) \ , 
\label{eq:twist31loop}
\ee
which gives $\gamma_{3,1} (1) = 8(1 - \log 2 )$, in obvious conflict 
with \eqref{eq:conjL}. However, as mentioned in \cite{LipatovStaudacher,Beccaria}, 
the twist-three expressions 
have been derived for physical, even values of $N$, and do not therefore need 
to be valid for unphysical, odd values of $N$. 
In fact, there is an important subtlety in the evaluation of 
$\gamma_L(N)$ at unphysical values of $N$ related to the two different 
prescriptions that exist in QCD for analytically continuing the harmonic 
sums entering the expansions of $\gamma_L(N)$ to generic values of the 
Mellin moment $N$. As discussed in \cite{KVanalytic}, 
there is a unique way to analytically continue sums with positive 
indeces. Sums with a negative index, however, such as $S_{-a,b,\ldots }$, 
have, due to their definition as an alternating series, a $(-1)^N$ factor. 
The oscillatory nature of this factor would, after analytical continuation, 
make the sums explode exponentially along the imaginary $N$ axis, and the 
inverse Mellin transforms would thus be ill-defined. Instead, if one 
chooses to analytically continue solely from even (or odd) values of $N$, 
the $(-1)^N$ factor can be set to a constant $+1$ (or $-1$), 
and well-behaved analytical continuations are obtained. 
The harmonic sums obtained by continuing from even $N$ are denoted $S^{(+)}$ 
(together with the corresponding indeces) and the sums obtained from negative 
values of $N$ are written $S^{(-)}$. It should be stressed that $S^{(+)}$ 
(respectively $S^{(-)}$) give incorrect 
values for odd (even) integer $N$. The two prescriptions then define two analytic 
expressions for the anomalous dimensions, $\gamma ^{(+)}(N)$ and $\gamma ^{(-)}(N)$. 

In QCD, both the positive and negative 
expressions are present, in the form of the singlet and non-singlet 
anomalous dimensions (see for instance \cite{BassoKorchemsky} for a 
recent discussion). In ${\cal N}=4$ supersymmetric Yang-Mills, 
however, physical states always correspond to even moments, 
and the $(+)$ prescription is therefore singled out. For example, in 
\cite{LipatovStaudacher} it was the $(+)$ prescription that was used to 
analytically continue the twist-two anomalous dimension obtained from the 
ABA to $N=-1$, where its singular behaviour could be 
compared to the predictions from BFKL on the leading singularity. At a 
pole, the singular behaviour of the two prescriptions, differ in sign. For example,
near $\omega \rightarrow 0$,
\be
S_{-a}^{(+)} (N + \omega) \sim \frac{(-1)^{N+1}}{\omega ^a} \ , \quad  
N = -1,\, -2,\,\ldots \ , \label{eq:plussing}
\ee
while
\be
S_{-a}^{(-)} (N + \omega) \sim \frac{(-1)^{N}}{\omega ^a} \ , \quad 
N = -1,\, -2,\,\ldots \label{eq:minussing}
\ee

The twist-three expressions show no oscillatory behaviour, and since they 
are extracted for even $N$, they may well be giving only the $(+)$ 
analytic continuation. It could then be argued that the twist-three dimension is 
written in terms of sums with positive indeces, for which the two 
prescriptions give the same result. However, the formulae in terms 
of positive indices could very well be an effective description 
only valid for even $N$. In contrast, in order to obtain the correct 
dispersion relation for twist-two, the $(-)$ prescription has to be 
used, since we are evaluating the harmonic sums at an odd value of $N$. 
We believe that in general, it is the $(-)$ prescription that should be used to 
test \eqref{eq:conjL}. 


\section{DGLAP and BFKL}

\begin{figure}[ht]
\begin{center}
\psfrag{Q}{$Q^2$} \psfrag{Qp}{$Q'^2$} \psfrag{k}{$k^2$} \psfrag{=}{$=$} \psfrag{+}{$+$} 
\includegraphics{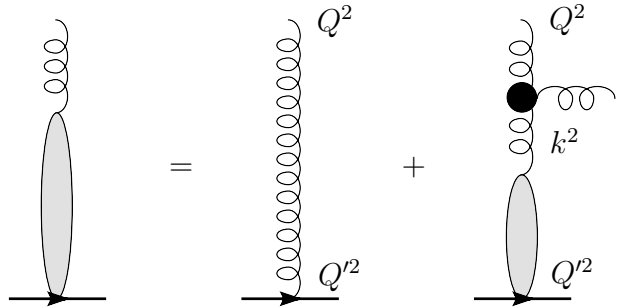}
\caption{\small The Bethe-Salpeter equation, performing the resummation 
of the logarithmic contributions to the evolution of the parton distribution 
functions in $Q^2$.} \label{fig:BS}
\end{center}
\end{figure}

\no
The Regge limit of high energy QCD corresponds to the scattering of two hadrons 
with the center of mass energy~$s$ 
much larger that the typical transverse scales, $Q^2$ and $Q'^2$. When $Q^2$ and $Q'^2$ are 
much larger than the QCD scale we can work in perturbation theory. In DIS $Q^2 \gg Q'^2$, 
with $Q^2$ the virtuality of the photon and $Q'^2$ the transversal scale of the target 
hadron. In this limit, the leading contribution to the evolution in $Q^2$ of the 
unintegrated parton distribution 
function $f(x,\, Q^2)$, which is related to the integrated parton 
distribution function $F(x,\, Q^2)$ through
\be
F(x,\, Q^2)=\int dk^2 f(x,\, k^2)\Theta(Q^2-k^2) \ ,
\ee
is determined by the Bethe-Salpeter integral equation
\be
f(x,\, Q^2)=f_{0}(x,\, Q^2) + 
2g^2 \int_{x}^{1}\frac{dz}{z}\int_{Q'^2}^{Q^2}\frac{dk^2}{Q^2}f\left(\frac{x}{z},\, k^2\right) \ , 
\label{eq:BS}
\ee
shown pictorially in figure \ref{fig:BS}. The Bethe-Salpeter equation takes this form 
for a kinematic region where we have, not only $x \ll 1$ and $z \ll 1$ corresponding 
to the Regge limit, but strict ordering in the longitudinal momenta, $z \gg x$, and 
also an ordering $Q^2 \gg k^2$ along the transversal momenta. 
For example, the assumption $z \ll 1$ implies that only the $1/x$ 
part of the gluon splitting function will be relevant, giving the 
$1/z$ factor in the integration kernel. 

Iterating the integral equation produces a sequence of ladder diagrams 
where the ordering in the transverse momenta leads to logarithms in 
the energy, $\log \left(\frac{s}{Q^2} \right)=\log \left(\frac{1}{x} \right)$, while the strict 
ordering of the transversal momenta produces the logarithmic 
collinear enhancement factors $\log \left(\frac{Q^2}{Q'^2}\right)$. 
Thus the previous Bethe-Salpeter integral equation is 
performing the perturbative resummation of double logarithms of the form
\be\label{1}
\left(g^2 \log \left(\frac{1}{x}\right)\log\left(\frac{Q^2}{Q'^2}\right)\right)^n \ .
\ee
Taking the Mellin transform with respect to both $x$ and $Q^2$, 
\footnote{If DGLAP conventions were used, the exponent of the 
$\left(\frac{Q^2}{Q'^2}\right)$-factor would simply be denoted 
$\gamma$. However, as later in this note we will relate $\gamma$ to 
the anomalous dimension, which we treat using spin chain conventions, the $-\gamma/2$ 
factor appears.}
\be
f(x,\, Q^2)=\int \frac{d\omega}{2\pi i} x^{-\omega}\int \frac{d\gamma}{2\pi i}
\frac{1}{Q^2}\left(\frac{Q^2}{Q'^2}\right)^{-\gamma/2} f(\omega,\, \gamma) \ , 
\label{eq:inverseMellin}
\ee
the solution to the Bethe-Salpeter integral equation is given by
\be
f(\omega,\, \gamma)=\frac{\omega f_{0}(\omega,\gamma)}{\omega + 4g^2 \frac{1}{\gamma}} \ ,
\ee
which leads to 
\be
f(x,\, Q^2) \sim \exp\left({\sqrt{8g^2 \log \left(\frac{1}{x}\right)
\log \left(\frac{Q^2}{Q'^2}\right)}}\right) \ ,
\ee 
corresponding to the resummation of the double logs (\ref{1}). In this collinear limit 
the DGLAP kernel in Mellin space is simply given by $-2/\gamma$. 

In contrast with the evolution in $Q^2$ that DGLAP gives, the BFKL equation 
provides us, in its domain of validity, with the behaviour of unintegrated 
parton distribution functions under changes of $x$ \cite{BFKL}. 
The kinematical regime where BFKL is defined corresponds to 
scattering of two hadronic objects with transversal 
scales of the same order. In these conditions we cannot impose strict 
ordering on the transversal momenta in 
the ladder diagrams, and we have resummations of single logarithms of type 
$\left(g^2 \log \left(\frac{1}{x}\right)\right)^n$. 
To a large extent, however, the full leading logarithmic (LLA) 
BFKL solution is 
reproduced by requiring that it 
gives the DIS $-2/\gamma$ pole in the limit $\gamma \rightarrow 0$, 
and by imposing symmetry under the exchange 
of the scales $Q$ and $Q'$. From \eqref{eq:inverseMellin} we see that, 
for fixed $x$, this corresponds to 
requiring invariance under $-2/\gamma \rightarrow (1+\gamma/2)$, which gives the pole 
\be
\frac{1}{\omega-2g^2\left(-\frac{2}{\gamma}+\frac{1}{1+\gamma/2} \right)} \ .
\ee
A complete analysis corrects the equation slightly in the region in between the 
two poles, and implies the LLA BFKL pole
\be
\frac{1}{\omega-2g^2\chi_{\hbox{\tiny{LLA}}}(\gamma)} \ ,
\ee
where
\be
\chi_{\hbox{\tiny{LLA}}}(\gamma)=2\psi(1)-\psi \left( - \frac{\gamma}{2} \right) 
- \psi \left( 1+ \frac {\gamma}{2} \right)
\ee
is called the BFKL kernel. Notice that when $\gamma \rightarrow 0$ 
the kernel $\chi_{\hbox{\tiny{LLA}}}(\gamma) \sim -2/\gamma$, 
in agreement with the DIS result.


\subsection{NLLA and scale dependence}

\no
In DIS the relevant scale $s_{0}$, relating $x$ and $s$ through $x=s_0/s$, 
is the photon virtuality $Q^2$. At LLA we do not have dependence on the scale $s_{0}$, 
but this situation changes when we go to next to leading logarithm 
approximation (NLLA) \cite{NLLA}. In particular if we are working in the 
BFKL regime the natural scale is the symmetric choice $s_{0}=QQ'$. As mentioned above, 
in the DIS regime we get contributions of the form 
$\left( g^2\log \left( \frac{s}{Q^2}\right)\log \left(\frac{Q^2}{Q'^2}\right)\right)^n$. 
Shifting to the symmetric scale $s_{0}=QQ'$ these lead to contributions with more 
collinear logarithms $\log \left(\frac{Q^2}{Q'^2}\right)$ than powers of $g^2$, 
producing non-physical singularities in the $\gamma \rightarrow 0$ limit such as 
$g^4/\gamma^3$. Also, from the renormalization group 
equations it follows immediately that there can not be more powers 
of $\log Q^2$ than powers of the coupling $g^2$. These double collinear 
logarithms, where the term ``double'' refers to the appearence of 
two logarithms for each power of the coupling, should therefore 
be cancelled by higher order corrections to the BFKL kernel. The most straightforward 
way to substract them is by introducing $\omega$ into the arguments of the 
digamma functions in the LLA BFKL kernel (see for instance \cite{Salam}, 
and references therein),
\be
\chi_{\hbox{\tiny{LLA}}}(\gamma)
\rightarrow 2\psi(1)-\psi\left( - \frac{\gamma}{2} 
+ \frac{\omega}{2}\right)-\psi\left(1+ \frac{\gamma}{2} + \frac{\omega}{2}\right) \ .
\ee
This shifted kernel coincides with the LLA kernel at lowest order, since $\omega$ starts 
at order $g^2$, and it resums large parts of the higher order contributions.

The shift in the digamma functions can be easily understood in terms of scale transformations 
of the Mellin transform. Writing out the scale $s_{0}$, the inverse Mellin transform 
\eqref{eq:inverseMellin} is given by
\be
f(x,\, Q^2)=\int \frac{d\omega}{2\pi i} \left(\frac{s}{s_{0}}\right)^{\omega}
\int \frac{d\gamma}{2\pi i}\frac{1}{Q^2}\left(\frac{Q^2}{Q'^2}\right)^{-\gamma/2} 
f(\omega,\, \gamma) \ .
\ee
It follows that a change of scale $s_{0} \rightarrow s_{0}\frac{Q'}{Q}$ corresponds to the shift
$-\gamma/2 \rightarrow -\gamma/2 + \omega/2$. In DIS we have a $-2/\gamma$ pole for 
small $\gamma$, when the scale is $Q^2$. This imples that the first non-constant 
digamma of the characteristic function should be $-\psi (-\gamma/2)$ at $s_0 = Q^2$, 
which implies that it shifts to $-\psi (-\gamma/2 + \omega /2)$ at $s_0 = QQ'$. 
Requiring symmetry between $Q$ and $Q'$, and therefore a $1/(1+\gamma/2)$ pole 
when $s_0 = Q'^2$, provides the argument of the last digamma function. 

Now let us recall that the DGLAP anomalous dimensions and 
their equivalent description in terms of dimensions 
of twist-two operators arise when studying the parton distribution functions in DIS. 
When comparing BFKL predictions with the anomalous dimensions 
obtained from the spin chain picture, we should therefore choose the asymmetric $Q^2$ scale. 
With that choice we get
\be\label{4}
\chi(\omega,\, \gamma) = 2\psi (1) - \psi 
\left( - \frac{\gamma}{2} \right) - \psi \left( 1 + \frac {\gamma}{2} + \omega \right) \ .
\ee


\subsection{The double logarithmic resummation}

\no
In what follows we will be interested not only in double logarithms of 
the type \eqref{1}, but also in purely non-collinear double logarithms, i.e., 
in contributions to the parton evolution where each power of the coupling $g^2$ is acompanied 
by two powers of $\big( \log s \big)$. Contrary to the case of the purely collinear 
double logarithms discussed in the previous subsection, these 
non-collinear double logarithms are not compensated for at 
higher orders in the perturbative expansion. 

One way to to resum the entire double logarithmic contribution 
to the parton evolution, including both $ \big( \log s \log Q^2 \big)$ and 
$\big( \log ^2 s \big)$ terms, is to modify the Bethe-Salpeter integral 
equation \eqref{eq:BS} by changing the kinematic region over which one integrates 
\cite{KotikovLipatov}. We still require that $z \gg x$, or 
equivalently $s \gg s'$, where $s' = \frac{Q^2}{z}$, but we 
now relax the ordering of the transverse momenta, moving in 
the direction of BFKL, allowing $k^2$ to be larger than $Q^2$, 
although still much smaller than $s$ or $s'$. Instead, we 
require that \footnote{The equation studied in \cite{KotikovLipatov} 
corresponded to a QED scattering amplitude, with a slighlty different 
structure than the case at hand, implying that the modification of the 
integration region performed was different than this one.}
\be
z \ll  \frac{Q^2}{k^2} \ ,
\ee
which is automatically satisfied if $k^2 < Q^2$ since $z \ll 1$,
 but becomes important in the extended kinematic region where $k^2 \gg Q^2$.
This additional condition mixes the transverse and longitudinal 
variables. As a result, collinear $\log Q^2$ 
logarithms can get substituted for additional logarithms in the energy. 
The way the double logarithmic contributions are generated from 
this change of kinematical region is shown in detail in appendix 
\ref{app:logs}. 

At the level of the Bethe-Salpeter kernel, the change of the integration region leads to a modification 
of the Mellin space kernel from $-2/\gamma$ to
\be
- \frac{2}{\gamma} + \frac{1}{\omega+\gamma/2} \ .
\ee
From this modified kernel we see that the pole
\be
\omega = 2g^2\left( - \frac{2}{\gamma} + \frac{1}{\omega+\gamma/2} \right) 
\ee
in the solution to the Bethe-Salpeter equation in Mellin space can be 
written \footnote{This is one of the two poles in $\gamma$. However, 
as noted in the appendix, the integration contour performed when 
taking the inverse Mellin transform only picks up one of the poles.}
\be\label{3}
\gamma = {\omega} \sqrt{1-\frac{8g^2}{\omega^2}} - \omega \ .
\ee


\subsection{BFKL anomalous dimensions: analytic continuation}

\no
In BFKL anomalous dimensions arise in a different way than for DGLAP. The solutions 
to the BFKL equation can be related to a four-point Green function of fields defined 
in impact parameter space. When the impact parameters of two of the fields get close, 
one can perform an operator product expansion where the anomalous dimensions
of the appearing operators are given by the BFKL kernel. One of the 
labels parameterizing the eigenfunctions of the BFKL 
equation is the conformal spin $n$. When $n=0$, the double logarithm corrected BFKL 
kernel is given by \eqref{4}. But in general, the BFKL anomalous dimensions 
depend on $n$, $\gamma=\gamma(\omega,\, n)$, and are given as solutions of
\be
\omega = 2g^2 \Big( 2\psi (1) - \psi \left( - \frac{\gamma}{2} \right) - 
\psi \left( 1 + \frac {\gamma}{2} + \omega + |n| \right) \Big) \ . 
\label{eq:BFKLDL} 
\ee
For ${\cal N}=4$ Yang-Mills it was suggested in \cite{KotikovLipatov} that by an analytic 
extension in $|n|$ we can get directly from BFKL the anomalous dimension of formal 
twist-two operators with negative spin. Defining $j = 1+|n|+\omega$, we are 
interested in moving in the $(\omega,\, |n|)$ plane to points with $|n|=-r-1$, 
where $r$ is a positive integer, and with $\omega$ going to zero as $-(r+1+|n|)$. 
Next, we should compare this double limit of $\gamma(\omega,\, |n|)$ with the 
analytic extension of the DGLAP anomalous dimensions for twist-two operators, 
$\gamma _2(N)$, analytically continued to $\gamma(-r+\omega)$ for $\omega \rightarrow 0$.

When we consider the analytic extension 
of DGLAP anomalous dimensions beyond one-loop we find 
terms of type $a_{i,\, r}g^{2i}/\omega^{2i-1}$, that for $i>1$ contain 
one more power of $g$ than powers of $\omega$. These are precisely of the 
form obtained when expanding the expression \eqref{3} for the double 
logarithm pole. The analytic extension of the anomalous dimensions 
thus contains a piece
\be
\gamma(-r+\omega)=\sum_{i} \frac {a_{i,\, r}}{\omega^{2i-1}}g^{2i}+ \cdots
\ee
which, invoking the relation to DIS, via BFKL, can be traced to the double 
logarithm contribution. 

Using the known perturbative results until four-loops 
(or three-loops if the ABA result is not trusted), 
one discovers the following relation between the double logarithm 
coefficients $a_{i,\, r}$ and the 
coefficients of the loop expansion of $\gamma(1)= \sum_{i}e_{i}g^{2i}$ at twist-two,
\be\label{two}
a_{i,\, r}=(-1)^{i}e_{i} \ ,
\ee
for even values of $r$. For odd values of $r$ we get $a_{2} = a_{3}=0$, which correspond to 
the typical behaviour of the BFKL pomeron. Assuming the previous relation holds to all-loops we 
observe that the contribution of the double logarithms to the anomalous dimension 
analytically extended to negative values of the spin for $r$ even is completely linked 
to the anomalous dimension $\gamma(1)$.

The double logarithm contribution can also be extracted, as is done in \cite{KotikovLipatov}, 
directly from the BFKL kernel. Approximating \eqref{eq:BFKLDL} by only keeping the 
singular parts of the poles at $\gamma = 0$ and $\gamma = -2 \omega$ one gets
\be
\omega = 2g^2 \left( - \frac{2}{\gamma } + \frac{1}{\omega + \gamma/2}\right) \ , \label{eq:BFKLDLpoles}
\ee
which simplifies to \eqref{3}. There is a subtlety in this derivation, however. For 
fixed coupling, when $\omega \rightarrow 0$, the $\gamma$ does not approach one of 
the poles, invalidating the pole approximation. \footnote{For $|n|=-1$, there is actually a solution 
of \eqref{eq:BFKLDL} where $\gamma$ approaches 0 as $\omega$ does. However, this 
solution $\gamma (\omega)$ does not seem to be related to the double logarithms.} 
This can be seen from equation \eqref{3} 
since it implies that $\gamma$ approaches an imaginary constant when $\omega$ 
tends to zero.  The solution is to let $g^2 \ll \omega$. The expression for the double logarithmic pole
is thus obtained from BFKL when $\omega$ is small, and the coupling is even smaller.


\subsection{Magnon dispersion relation and double logarithms}

\no
As discussed above the spin chain representation of the anomalous dimensions 
suggests to interpret $\gamma(1)$ as the energy for a magnon with the minimal 
non-vanishing momentum in a chain of length-two with periodic boundary conditions. 
This interpretation of $\gamma(1)$, together with (\ref{two}), leads to
\be
\gamma^{(DL)}(-r+\omega)=\omega 
E \left( p=\pi,\, g \rightarrow \frac{ig}{\omega} \right) \ ,
\ee
where by $\gamma^{(DL)}$ we mean the double logarithm contribution to the anomalous dimension 
and where, as before, we assume $r$ even. Once we have related the double logarithm 
contribution to the magnon energy, we can use the 
information about its contribution in DGLAP to determine at all-loops the form of the 
magnon dispersion relation, $E(ig/\omega)$. The logic flow of the discussion 
here is first to interpret $\gamma(1)$ as the single magnon energy, secondly to relate 
$\gamma(1)$ with the double logarithm contribution and finally to get the form of the 
magnon energy from the DGLAP kernel including the double logarithm pieces. 
As discussed above, the double 
logarithm contribution to $\gamma$ is given by (\ref{3}), and therefore we get
\be
E \Big(p=\pi,\, g \rightarrow \frac{ig}{\omega} \Big) = 
\omega \sqrt{1-\frac{8g^2}{\omega^2}} - \omega \ ,
\ee
in agreement with the ABA prescription.

In addition, this agreement gives added weight to the currently 
used form of the ${\cal N}=4$ dispersion relation. The algebraic 
contruction of the ABA \cite{BeisertS} 
introduces a dispersion relation of the form \eqref{eq:dispersion}. 
However, there is nothing that prevents the algebraically introduced coupling constant 
from being an arbitrary function of the physical coupling~$g$.~\footnote{See for instance 
the related recent proposal in \cite{GromovVieira}.}

In extracting the dispersion relation from the double logarithmic 
approximation of BFKL we did not assume that the magnon itself had an 
interpretation in this formalism. However, we believe that there is a 
BFKL magnon candidate. The solution to the Bethe-Salpeter equation corresponding 
to the double logarithmic approximation can be related to a 
certain $t$-channel partial wave expansion (see appendix D in 
\cite{KotikovLipatov}). The amplitude for such a partial 
wave is given by (equation (D2) in \cite{KotikovLipatov}) 
\be
f_\omega = \frac{\omega ^2}{4g^2}\left( 1 - \sqrt{1-\frac{8g^2}{\omega ^2}} \right) \ .
\ee
We can therefore speculate that the relation between the spin chain magnon and 
BFKL is as presented in table \ref{tab:magBFKL}. A single magnon is thus 
identified with a partial-wave in the double logarithmic approximation. 
Including subleading terms in the integral equation would then correspond 
to adding interactions between magnons.

\begin{table}[ht]
\begin{center}
		\begin{tabular}{ll}
			{\bf Spin chain} & {\bf BFKL} \\
			\hline
			Magnon     & Partial wave in double logarithmic approximation \\
			E  &  Partial wave amplitude (re-scaled)  \\
			$\sin \left( \frac{p}{2} \right)$ &  $i/\omega$ \\
			$g$        &  $g$ \\
		\end{tabular}
		\caption{\small BFKL description of the spin chain magnon.}
\label{tab:magBFKL}
\end{center}
\end{table}	

In fact, this relationship is entirely analogous to the 
approach in \cite{FaddeevKorchemsky}  
linking high energy QCD and the XXX$_{s=0}$ spin chain. 
Eigenfunctions of the Bethe-Salpeter kernel, which amount to 
partial waves in that case, where mapped to magnons of the spin chain, and 
the spin chain hamiltonian was obtained. The spin $0$ 
construction is, however, limited to leading order. Here we 
have possibly the starting point for a map from all-order BFKL 
to a spin chain. However, obtaining the explicit map may be difficult, because 
it would entail constructing the complete all-loop dilatation 
operator, including wrapping effects. Still, a partial map could 
shed light on both BFKL and the ${\cal N}=4$ spin chain.


\section{Conclusions}

\no
In this note we have put forth a series of conjectures, based on perturbative 
evidence, on the relation of the dispersion relation for planar 
${\cal N}=4$ Yang-Mills to the 
double logarithmic contributions to the anomalous dimension for 
twist-two operators. Let us briefly recall them: 1. The first 
conjecture relates the perturbative coefficients $e_i$
in the coupling for the anomalous dimension of twist-two operators, $\gamma_2(N)$, 
at $N=1$, to the coefficients for the double logarithm contributions to 
$\gamma_2(N)$ at $N=-r$, for even values of $r$. We have presented evidence that 
$e_i = (-1)^i a_{i , r}$. 2. Secondly, we have suggested that the 
anomalous dimension $\gamma_2(1)$, evaluated 
using the $(-)$ analytic extension for the harmonic sums, corresponds 
to the dispersion relation for a single magnon of momentum $p=\pi$, 
$\gamma_2(1) = E(p=\pi)$. 3. Our last statement is an extension to 
twist-$L$ operators, $\gamma_L(1) = E(p=\pi/L)$, whenever 
the $(-)$ analytic extension is defined.

The first conjecture is on firmest footing since it  
seems that there is some principle restricting the possible harmonic sums 
which enter the perturbative expansions of the anomalous dimensions, so that 
their evaluations at $N=1$ and at negative, even integers, are indeed related. 
Furthermore, only the double logarithm contribution is matched to $\gamma _2(1)$. 
That is, terms that are subleading in either the coupling, or in $1/\omega$, 
in the expansion of the anomalous dimensions around $-r$, for $r$ even, do 
not enter in the anomalous dimension at $N=1$. This is a highly non-trivial statement, 
since at $N=1$ all harmonic sums contribute to the anomalous dimension, 
while only the most singular sums contribute to the double logarithm expansion. 
Notably, nested harmonic sums typically do not affect the double 
logarithms. 
  
One might then wonder whether wrapping effects could spoil 
the validity of the first two conjectures. Wrapping is understood as responsible 
for the mismatch for twist-two operators between the ABA and BFKL at 
four-loops \cite{LipatovStaudacher}. From the viewpoint of BFKL, 
wrapping is never an issue and must automatically be included in 
the BFKL answer. Since the double logarithmic contribution is, at 
weak-coupling and at all-loops, determined by 
equation \eqref{3}, and one could in principle derive also $\gamma_2 (1)$ 
to all orders solely from BFKL \cite{KotikovLipatov}, there is no reason 
to believe that something special will happen at fourth loop order that 
ruins conjecture 1. If one then invokes the intuitive idea of 
$\gamma _2(1)$ giving the energy of a single magnon as 
justification for the second conjecture, one is lead to the 
conclusion that wrapping effects should not modify the single magnon dispersion 
relation at weak-coupling (strong-coupling is, of course, 
an entirely different issue). Most likely, before including wrapping effects, the ABA answer 
should be consistent with the relation $\gamma_2(1)=E(p =\pi)$, implying that
wrapping modifications to the ABA should correspond to combinations of 
harmonic sums respecting transcendentality, and vanishing when 
evaluated at $N=1$. This is in fact the case for the ad hoc proposal 
in \cite{LipatovStaudacher}. However an important problem that 
we have not considered in this note is if a potential extension of 
BFKL to strong 't~Hooft coupling (see for instance
\cite{Polchinski}) would modify the form of the double logarithmic 
contribution. 

The third conjecture, by contrast, is merely a wild idea based on the 
intuitive notion underlying the second conjecture. It could very well be 
that it is only valid at even $L$, or that the relevant magnon momentum 
is not the minimal $2\pi /L$, but something else. 

We have also, by analogy with the emergence of one-loop integrability in high 
energy QCD, conjectured 
that the all-loop magnon appears in BFKL in the form of a partial wave in 
the double logarithmic approximation. It would be very interesting if this 
correspondence could be extended to a complete BFKL--spin chain map.

As a final comment, let us recall that the magnon dispersion relation for planar 
${\cal N}~=~4$ Yang-Mills is intimately related to the string BMN formula and 
moreover it can be derived, barring possible differences between the 
algebraically introduced coupling and the physical coupling, 
from the centrally extended symmetry algebra \cite{BeisertS}. 
It would be extremely interesting to find glints of these 
structures in the double logarithmic contributions to the Bethe-Salpeter 
equations governing the parton distribution functions. Perhaps one could use this information to extend the BFKL - spin chain map beyond the single magnon.


\vspace{5mm}
\centerline{\bf Acknowledgments}

We are grateful to R. A. Janik and A. Sabio Vera for comments and discussions. 
C.~G. would like to thank the Galileo Galilei Institute for hospitality and the 
INFN for partial support during the completion of this work. 
The work of J. G. is supported by a Spanish FPU grant and by a European 
fellowship through MRTN-CT-2004-005104. This work is also partially 
supported by the Spanish DGI under contracts FPA2003-02877 and FPA2003-04597 
and by the CAM project HEPHACOS P-ESP-00346.


\appendix

\section{The resummation of double logarithms from the integral equation}

\label{app:logs}
We will now show how a modification of the kinematic region used in the 
Bethe-Salpeter equation
\be
f(x,\, Q^2)=f_{0}(x,\, Q^2) + 
2g^2 \int_{x}^{1}\frac{dz}{z}\int_{Q'^2}^{Q^2}\frac{dk^2}{Q^2}f\left(\frac{x}{z},\, k^2\right) 
\ , \label{eq:BS1}
\ee
can produce all types of double logarithmic terms that are consistent with 
renormalization group constraints, by which we mean that the number of 
collinear logarithms are not allowed to exceed the order in perturbation 
theory. Firstly, we will drop the requirement of transversal ordering $Q^2 \gg k^2$. 
This changes the upper integration limit in the integral over transverse 
momenta from $Q^2$ to $s = \frac{Q^2}{x}$. Secondly, we add the condition that
\be
z  \ll \frac{Q^2}{k^2} \ ,
\ee
which is not trivially satisfied when $k^2 > Q^2$. This causes the upper limit of the integral 
over $z$ to become the smaller of $1$ or $\frac{Q^2}{k^2}$. Therefore \eqref{eq:BS1} is modified to
\be
f(x,\, Q^2)=f_{0}(x,\, Q^2) + 
2g^2 \int_{Q'^2}^{Q^2/x}\frac{dk^2}{Q^2} \int_{x}^{\min \left(1,\, 
Q^2/k^2\right)}\frac{dz}{z} f\left(\frac{x}{z},\, k^2\right) \ . \label{eq:BS2}
\ee
For the two different cases present in the integration limit $\min \left(1,\, 
\frac{Q^2}{k^2}\right)$, the integration over transverse momenta comes from 
different regions ($k^2 > Q^2$ or $k^2<Q^2$, respectively) and we get
\be
f(x,\, Q^2)=f_{0}(x,\, Q^2) + 
2g^2 \int_{Q'^2}^{Q^2}\frac{dk^2}{Q^2} \int_{x}^{1}\frac{dz}{z} f\left(\frac{x}{z},\, 
k^2\right) + 2g^2 \int_{Q'^2}^{Q^2/x}\frac{dk^2}{Q^2} \int_{x}^{Q^2/k^2}\frac{dz}{z} 
f\left(\frac{x}{z},\, k^2\right) \ . \label{eq:BS3}
\ee
By iteration this equation produces the perturbative expansion of the double logarithmic terms. 
For example, with the simplest possible initial distribution $f_0(x\, Q^2)=\frac{1}{Q^2}$,
 where the factor $1/Q^2$ has to appear since the integrated parton distribution should be 
dimensionless, one obtains
\begin{align}
f(x,\, Q^2)&= \frac{1}{Q^2} + \frac{2g^2}{Q^2}\left( \log  \frac{Q^2}{Q'^2}  \log \frac{1}{x} 
+ \frac{1}{2}\log ^2 \frac{1}{x}  \right) + \nonumber \\
 &+ \frac{4g^4}{Q^2}\left(\frac{1}{4}\log ^2 \frac{Q^2}{Q'^2}\log ^2  \frac{1}{x}    
+ \frac{1}{3}\log \frac{Q^2}{Q'^2}\log ^3 \frac{1}{x} +  \frac{1}{12}\log ^4  \frac{1}{x} \right) 
+ {\cal O}(g^6) \ . \label{eq:iteration}
\end{align}
Iteration of the first integral in \eqref{eq:BS3}, which is the same integral 
as in \eqref{eq:BS1}, produces double logarithms of the form $\left( g^2 \log  
\frac{Q^2}{Q'^2} \log \frac{1}{x} \right)^n$, while iteration of the second 
integral produces double logarithms in the energy $\left( g^2 \log ^2 \frac{1}{x} \right)^n$. 
Combining the two terms when iterating leads to mixed cases.

However, usually one introduces \eqref{eq:BS3} because the double logarithmic 
contribution makes the perturbation expansion badly divergent, such as is the 
case when the energy is so large that $g^2 \log ^2 \frac{s}{Q^2} $ is of order 
unity or larger. Solving the integral equation provides a resummation to all 
orders of the double logarithms. This can be done by Mellin transforming the 
distributions,
\be
f(x,\, Q^2)=\int ^{\sigma + i\infty}_{\sigma - i\infty } \frac{d\omega}{2\pi i} 
x^{-\omega}\int ^{\sigma ' + i\infty}_{\sigma ' - i\infty } \frac{d\gamma}{2\pi i}
\frac{1}{Q^2}\left(\frac{Q^2}{Q'^2}\right)^{\gamma} f(\omega,\, \gamma) \ , \label{eq:inverseMellin2}
\ee
where the integration contour for the $\gamma$ integral runs parallel to the 
imaginary axis with a positive real part, $\sigma ' > 0$, and the $\omega$ 
integration contour is also parallel to the imaginary axis with $\sigma - \sigma ' > 0$. 
As they are much more convenient in performing the following calculations, we are 
using DGLAP conventions in this appendix for $\gamma$ as oposed to the spin chain conventions 
used in the main text. The results obtained can be translated to the spin chain 
conventions by simply letting
\be
\gamma \rightarrow -\frac{\gamma}{2} \ .
\ee

If one introduces \eqref{eq:inverseMellin2} into \eqref{eq:BS3}, and performs 
the integrals over $z$ and $k^2$, the first integral becomes 
$f(\omega,\, \gamma)/\omega \gamma$, 
while the second integral transforms to $f(\omega,\, \gamma)/\omega(\omega - \gamma)$, 
which gives the double logaritmic pole
\be
f(\omega,\, \gamma) \sim \frac{1}{\omega - 2g^2 \left[ \frac{1}{\gamma} 
+ \frac{1}{\omega - \gamma} \right]} \ . 
\ee
The double Mellin transform is full of subtleties, however, and the correct 
answer is not obtained simply by introducing \eqref{eq:inverseMellin2} 
into \eqref{eq:BS3}. In appendix D of \cite{KotikovLipatov} an alternative 
method is used to pass to Mellin space when solving a similar integral 
equation. For the simple initial distribution $f_0(x,\, Q^2) = 1/Q^2$ one obtains 
\be
f(\omega ,\, \gamma) = \frac{(\omega - 2\gamma)\gamma \omega 
(2g^2)^{-2}}{\omega - 2g^2 \left[ \frac{1}{\gamma} + \frac{1}{\omega - \gamma}  \right]} \ .
\label{eq:Mellinsol}
\ee
One can now transform back to the physical variables $x$ and $Q^2$ by 
performing the integrals in \eqref{eq:inverseMellin2}. We can re-write \eqref{eq:Mellinsol} as
\begin{align}
f(\omega,\, \gamma) &= \frac{(\omega - 2\gamma)(\gamma - \omega)\gamma ^2 
(2g^2)^{-2}}{\gamma ^2 - \omega \gamma + \lambda}   \nonumber \\
&=\frac{(\omega - 2\gamma)(\gamma - \omega)\gamma ^2 (2g^2)^{-2}}{ 
\left( \gamma - \frac{1}{2}\left( \omega + \sqrt{\omega ^2 - 8g^2} 
\right) \right)\left( \gamma - \frac{1}{2}\left( \omega - 
\sqrt{\omega ^2 - 8g^2} \right) \right)} \  . \label{eq:Mellinsol2}
\end{align}
Now, performing the integral over $\gamma$ we will only pick up the pole at 
\be
\gamma = \frac{1}{2}\left( \omega - \sqrt{\omega ^2 - 8g^2} \right) \ , \label{eq:polegamma}
\ee
since the two poles lie on either side of the $\gamma$ contour, 
and $\left( \frac{Q^2}{Q'^2} \right) > 1$ implies that we must close the contour towards the left. 

After having performed the $\gamma$ integral we are left with
\begin{eqnarray}
&& Q^2 f(x,\, Q^2) =  \nonumber \\
&& = \int \frac{d\omega}{2\pi i}x^{-\omega}\frac{\sqrt{\omega ^2-8g^2}
\left( \omega + \sqrt{\omega ^2-8g^2} \right)\left( \frac{\omega}{2} 
- \frac{1}{2}\sqrt{\omega ^2-8g^2} \right)^2}{8g^4 \sqrt{\omega ^2-8g^2}}
\left( \frac{Q^2}{Q'^2} \right)^{\frac{1}{2}\left( \omega - \sqrt{\omega ^2 - 8g^2} \right)} \nonumber \\
&& = \int \frac{d\omega}{2\pi i}x^{-\omega}\frac{1}{4g^2}
\left( 1 - \sqrt{1 - \frac{8g^2}{\omega ^2}} \right)  
\exp \left[\frac{\omega}{2}\left( 1 - \sqrt{1 - \frac{8g^2}{\omega ^2}} 
\right)\log  \frac{Q^2}{Q'^2}\right] \ .
\end{eqnarray}
This integral can be evaluated, for example by performing a saddle 
point approximation. Instead, let us simply note how the perturbative 
expansion of this expression consists only of double logarithms. 
The inverse Mellin transform of $1/\omega ^{r+1}$ 
is $\left( \frac{1}{r!}\log ^r \frac{1}{x} \right)$, and each instance of the 
coupling $g^2$ is acompanied by either $1/\omega ^2$ 
or by $\left( \frac{1}{\omega} \log  \frac{Q^2}{Q'^2}\right)$, explaining the 
double logarithms. Also, at most one factor of 
$\log \left( \frac{Q^2}{Q'^2} \right) $ can appear at 
each order in perturbation theory, which must be the case 
in order for the double logarithmic approximation to be compatible 
with the renormalization group equations.



\end{document}